\documentclass[useAMS,usenatbib]{mn2e}

\usepackage{amsmath}
\usepackage{amssymb}
\usepackage{graphicx}
\usepackage{txfonts}
\usepackage{natbib}
\graphicspath {{./figures/}}
\newcommand{\dd}{\mathrm{d}}
\newcommand{\e}{\mathrm{e}}

\newcommand{\aap}{A\&A}

\newcommand{\ao}{Appl. Opt.}
\newcommand{\apj}{ApJ}
\newcommand{\apjs}{ApJS}
\newcommand{\apjl}{ApJL}
\newcommand{\araa}{ARA\&A}
\newcommand{\mnras}{MNRAS}
\newcommand{\pasp}{Publi. Astron. Soc. Pac.}
\newcommand{\prd}{Phys. Rev. D}
\newcommand{\jcap}{J. Cosmology Astroparticle Phys.}

\title[Modelling excesses of massive galaxy clusters]{On the modelling of the excesses of galaxy clusters over high-mass thresholds}

\author[J.-C. Waizmann, S. Ettori and L. Moscardini]{J.-C. Waizmann$^{1,2,3}$\thanks{E-mail:
jcwaizmann@oabo.inaf.it}, S. Ettori$^{2,3}$ and L. Moscardini$^{1,2,3}$\\
$^{1}$Dipartimento di Astronomia, Universit\`{a} di Bologna, via Ranzani 1, 40127 Bologna, Italy\\
$^{2}$INAF - Osservatorio Astronomico di Bologna, via Ranzani 1, 40127 Bologna, Italy\\
$^{3}$INFN, Sezione di Bologna, viale Berti Pichat 6/2, 40127 Bologna, Italy}

\begin{document}

\date{Accepted 2012 March 2. Received 2012 February 28; in original form 2012 January 17}

\pagerange{\pageref{firstpage}--\pageref{lastpage}} \pubyear{2012}

\maketitle

\label{firstpage}

\begin{abstract}
In this work we present for the first time an application of the Pareto approach to the modelling of the excesses of galaxy clusters over high-mass thresholds. The distribution of those excesses can be described by the generalized Pareto distribution (GPD), which is closely related to the generalized extreme value (GEV) distribution. After introducing the formalism, we study the impact of different thresholds and redshift ranges on the distributions, as well as the influence of the survey area on the mean excess above a given mass threshold. We also show that both the GPD and the GEV approach lead to identical results for rare, thus high-mass and high-redshift, clusters. As an example, we apply the Pareto approach to ACT-CL J0102--4915 and SPT-CL J2106--5844 and derive the respective cumulative distribution functions of the exceedance over different mass thresholds. We also study the possibility to use the GPD as a cosmological probe. Since in the maximum likelihood estimation of the distribution parameters all the information from clusters above the mass threshold is used, the GPD might offer an interesting alternative to GEV-based methods that use only the maxima in patches. When comparing the accuracy with which the parameters can be estimated, it turns out that the patch-based modelling of maxima is superior to the Pareto approach. In an ideal case, the GEV approach is capable to estimate the location parameter with a percent level precision for less than $ \sim 100 $ patches. This result makes the GEV based approach potentially also interesting for cluster surveys with a smaller area.
\end{abstract}


%------------------------------------------
\section{Introduction}\label{sec:intro}
%------------------------------------------
Extreme value statistics (EVS), pioneered by the works of \cite{Fisher1928} and \cite{Gnedenko1943}, is a branch of statistics that deals with the statistical modelling of extreme events that substantially deviate from the mean behaviour. The principal characteristics of EVS is the fact that, for independently identically distributed (i.i.d.) random variables, the distribution of the extrema converges to a member of the generalized extreme value (GEV) distribution.

While EVS being widely spread in the environmental and economic sciences, it has not seen many applications to astrophysics so far. For the first time, EVS was applied to the study of the statistics of the brightest cluster galaxies by \cite{Bhavsar1985} and subsequently to the temperature maxima in the CMB by \cite{Coles1988}. It has also been applied to the solar cycle \citep{Asensio2007} and to solar radio bursts \citep{Rosa2010} and in a cosmological context to Gaussian random fields \citep{Colombi2011}.

Recently, mainly triggered by the detection of very massive galaxy clusters at high redshifts like XMMU J2235.3−2557 at $z=1.4$ \citep{Mullis2005,Rosati2009, Jee2009}, ACT-CL J0102 at $z=0.87$ \citep{Marriage2011, Menanteau2011} and SPT-CL J2106 at $z=1.132$ \citep{Foley2011, Williamson2011}, the application of EVS on massive clusters has been studied in several works. \cite{Davis2011} related for the first time the GEV distribution parameters to cosmological quantities and compared the approach to numerical N-body simulations. The impact of primordial non-Gaussianity on the EVS of galaxy clusters has been studied by \cite{Chongchitnan2011}. A direct approach, based on the exact rather than the asymptotic form, has been utilized by \cite{Harrison&Coles2011,Harrison&Coles2011b} to study the halo mass function and the possibility to use extreme clusters to test cosmological models. \cite{Waizmann2011a} proposed to utilise the GEV distribution as a cosmological probe by dividing the survey area in small equally sized patches allowing to reconstruct the distribution of the most massive haloes in those patches. GEV was also used in \cite{Waizmann2011b} to show that none of the known massive clusters alone is in conflict with $\Lambda$CDM and that there is no indication of high-$ z $ clusters to be rarer than low-$ z $ ones.

In this work we study the application of an alternative approach in the framework of EVS, which is based on the statistical modelling of the distribution of excesses (hereafter referred also to as exceedances) over a high threshold. It has been shown \citep{Pickands1975, Balkema1974} that for high thresholds the distribution of exceedances converges to the generalized Pareto distribution (GPD) which is closely related to the GEV distribution. The GPD distribution function allows to infer the probability that a given observation exceeds a high threshold by a certain amount and hence we utilise this approach to derive the distribution of the exceedances in mass of galaxy clusters over a high-mass threshold. Since surveys based on the Sunyaev-Zeldovich (SZ)-effect \citep{Sunyaev1972, Sunyaev1980}, like the \textit{South Pole Telescope} (\textit{SPT}) survey \citep{Carlstrom2011} for instance, can be considered to be mass-limited, we study as well whether a GPD based approach could be utilized as a cosmological probe.

This paper is structured according to the following scheme. In Section~\ref{sec:EVS} we introduce extreme value statistics by discussing the application of the Gnedenko approach to massive galaxy clusters in Section~\ref{sec:GEV}. This is followed by an introduction to the modelling of exceedances with the Pareto approach and its connection to the Gnedenko approach in Section~\ref{sec:GPD}. In Section~\ref{sec:exceed}, we apply the Pareto approach to massive high-$ z $ clusters in general. This is followed by an example application to two observed clusters in Section~\ref{sec:example_appl}, where in Section~\ref{subsec:prep_consid} we discuss several observational effects that have to be taken into account and in Section~\ref{subsec:single_results} we present the results of this exercise. After this analysis, we discuss in Section~\ref{sec:param_estim} the possible application of exceedance models to SZ cluster surveys as a cosmological probe and compare it with a GEV based approach. Then, we summarise our findings in the conclusions in Section~\ref{sec:conclusions}.     
%-------------------------------------
\begin{figure}
\centering
\includegraphics[width=0.9\linewidth]{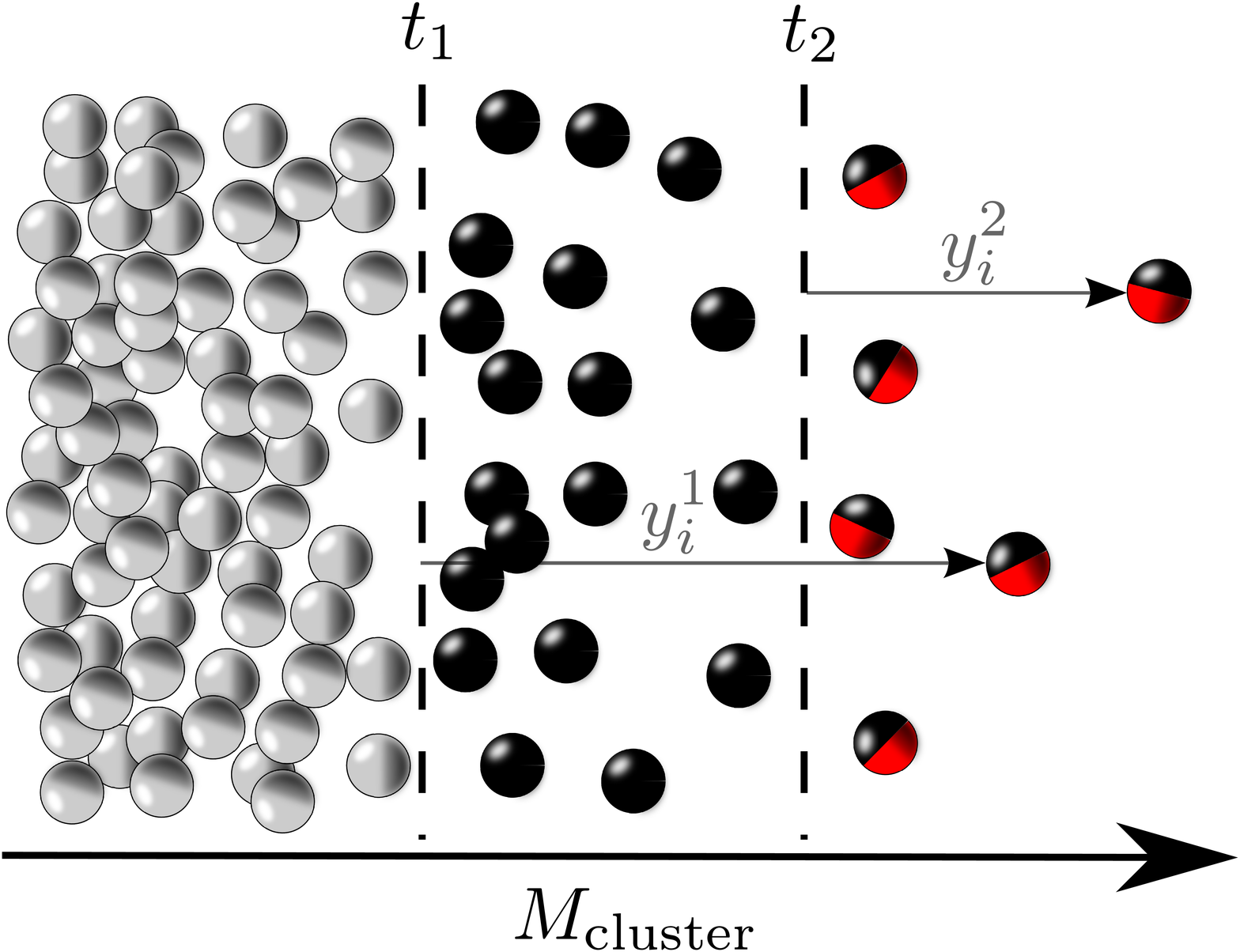}
\caption{Illustrative scheme of the exceedance model. Here $t_1$ and $t_2$ illustrate two different thresholds in cluster mass and $y^1_i$, $y^2_i$ are the corresponding exceedances over the respective threshold for an arbitrary cluster with tag $i$.}\label{fig:scheme}
\end{figure}
%-------------------------------------
\section{Extreme value statistics}\label{sec:EVS}
%-------------------------------------
Extreme value statistics (for an introduction see e.g. \cite{Gumbel1958, Kotz2000, Coles2001, Reiss2007}) concerns with the stochastic behaviour of the extremes (in what follows we consider only maxima) of i.i.d. random variables. In this sense EVS tries to model the unlikely and to give a quantitative answer to the question how frequent unusual observations are. There are two different approaches that are utilized in the literature and that will be briefly summarized in the following.
%------------------------------------------------------
\subsection{The Gnedenko approach}\label{sec:GEV}
%------------------------------------------------------
The Gnedenko approach deals with the modelling of the block maxima $ M_n $ of i.i.d. random variables $X_i$, which are defined as
\begin{equation}
M_n=\max(X_1,\dotsc X_n).
\label{eq:blockmax}
\end{equation}
It has been shown \citep{Fisher1928, Gnedenko1943} that for $n\rightarrow\infty$ the limiting cumulative distribution function (CDF) of the renormalised block maxima is given by one of the extreme value families: Gumbel (Type I), Fr\'{e}chet (Type II) or Weibull (Type III). As independently shown by \cite{vonmises1954} and \cite{Jenkinson1955}, these three families can be unified as a general extreme value distribution (GEV)
\begin{equation}\label{eq:GEV}
  F_{\rm GEV}\left(x;\alpha,\beta,\gamma\right) =\e ^{q(x)},
\end{equation}
with
\begin{equation}\label{eq:GEV_detail}
 q(x) = \left\{ 
  \begin{array}{l l}
     -\left[1+\gamma \left(\frac{x-\alpha}{\beta}\right)\right]^{-1/\gamma},& \; {\rm for}\;\gamma\neq 0,\\
    \e^{-\left(x-\alpha\right)/\beta},& \; {\rm for}\;\gamma = 0,\\
  \end{array} \right.
\end{equation}
with the location, scale and shape parameters $\alpha$, $\beta$ and $\gamma$. In this generalisation, $\gamma=0$ corresponds to the Type I,  $\gamma>0$ to Type II and $\gamma<0$ to the Type III distributions. The corresponding probability density function (PDF) is given by
\begin{equation}
f_{\rm GEV}\left(x;\alpha,\beta,\gamma\right)=\frac{\dd F_{\rm GEV}\left(x;\alpha,\beta,\gamma\right)}{\dd x}.
\end{equation}
From now on we will adopt the convention that capital initial letters denote the CDF (like $F_{\rm GEV}\left(x;\alpha,\beta,\gamma\right)$) and small initial letters denote the PDF (like $f_{\rm GEV}\left(x;\alpha,\beta,\gamma\right)$).

A formalism for the application of GEV to the most massive galaxy clusters has been introduced by \cite{Davis2011} and is briefly summarized in the following. By introducing the random variable $u=\log_{10}(m)$, the CDF of the most massive halo reads
\begin{equation}
{\rm Pr}\lbrace u_{\rm max} \le u\rbrace
\equiv \int_0^{u}p(u_{\rm max}) \,\dd u_{\rm max}.
\label{eq:cdf}
\end{equation}
This probability has to be equal to the one of finding no halo with a mass larger than $u$. On scales  ($\geq 100\,\textrm{Mpc/h}$) for which the clustering between galaxy clusters can be neglected, the CDF is given by the Poisson distribution for the case of zero occurrence \citep{Davis2011}:
\begin{equation}
P_0(u)=\frac{\lambda^k \e^{-\lambda}}{k!}=\e^{-n_{\rm eff}( > u) V},
\end{equation}
where $n_{\rm eff}( > u)$ is the effective comoving number density of halos above mass $u=\log_{10}(m)$ obtained by averaging and $V$ is the comoving volume. By assuming that equation~\eqref{eq:cdf} can be modelled by $F_{\rm GEV}\left(u;\alpha,\beta,\gamma\right)$, it is possible to relate the GEV parameters to cosmological quantities by Taylor-expanding both $F_{\rm GEV}\left(u;\alpha,\beta,\gamma\right)$ and $P_0(u)$ around the peaks of the corresponding PDFs. By comparing the individual first two expansion terms with each other, one finds \citep{Davis2011} 
\begin{eqnarray}\label{eq:parameters}
\gamma = n_{\rm eff}(>m_0)V-1, \quad \beta =
\frac{(1+\gamma)^{(1+\gamma)}}{\left.\frac{\dd\,n_{\rm eff}}{\dd\,m}\right|_{m_0}Vm_0\ln 10}, \nonumber \\
\alpha = \log_{10} m_0 - \frac{\beta}{\gamma}[(1+\gamma)^{-\gamma} -1],
\end{eqnarray}
where $m_0$ is the most likely maximum mass and $\left.\dd\,n_{\rm eff} / \dd\,m \right|_{m_0}$ is the effective mass function evaluated at $m_0$ which relates to the effective number density $n_{\rm eff}(>m)$ via
 \begin{equation}
 \left.\frac{\dd\,n_{\rm eff}}{\dd\,m}\right|_{m_0}=- \left.\frac{\dd\,n_{\rm eff}(>m)}{\dd\,m}\right|_{m_0}.
\end{equation}
 The most likely mass, $m_0$, can be found \citep{Davis2011} by performing a root search on
\begin{equation}\label{eq:m0_num}
 \left.\frac{\dd\,n_{\rm eff}}{\dd\,m}\right|_{m_0}+m_0\left.\frac{\dd ^2\,n_{\rm eff}}{\dd\,m^2}\right|_{m_0}+m_0V\left(\left.\frac{\dd\,n_{\rm eff}}{\dd\,m}\right|_{m_0}\right)^2=0.
\end{equation}
For calculating $n_{\rm eff}$ we utilized the mass function introduced by \cite{Tinker2008} and fix the cosmology to $(h,\Omega_{\Lambda0},\Omega_{\rm m0},\sigma_8)=(0.7,0.73,027,0.81)$ based on the \textit{Wilkinson Microwave Anisotropy} 7-yr (\textit{WMAP}7) results \citep{Komatsu2011}.
%------------------------------------------------------
\subsection{The Pareto approach}\label{sec:GPD}
\begin{figure}
\centering
\includegraphics[width=0.95\linewidth]{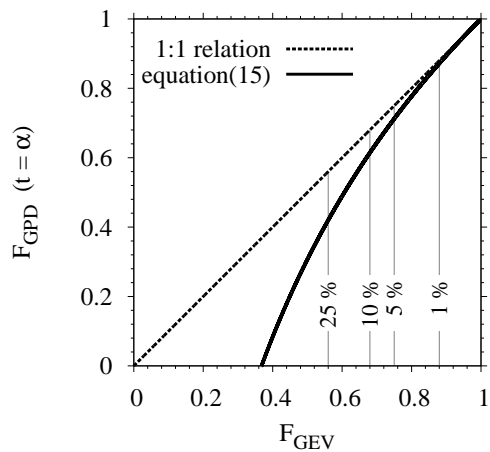}
\caption{Relation between the CDFs of the GEV ($ x $-axis) and the GPD ($ y $-axis) approach under the assumption of $ t=\alpha $. The dotted line illustrates the $ 1:1 $ relation and the solid line depicts the true relation according to equation~\eqref{eq:gpd_gev_relation}. The vertical lines show the deviations from the $ 1:1 $ relation for different values of the CDF based on the GEV approach.}\label{fig:gpdvsgev}
\end{figure}

%------------------------------------------------------
Exceedance theory spreads, originating mainly from hydrological literature (see e.g. \cite{Fitzgerald1989, Balkema1974}), into many fields of applied statistics. The basic notion is that, instead of studying the distribution of the maxima in blocks of random variables as in equation~\eqref{eq:blockmax}, an alternative view is to consider realisations $X_i$ drawn from an underlying distribution $F$ as extreme, if they exceed a very high threshold $t$ as depicted in Fig.~\ref{fig:scheme}. Thus, one is interested in the conditional probability
\begin{equation}
{\rm Pr}\lbrace X>t+y\;|\;X>t\rbrace = \frac{1-F(t+y)}{1-F(t)},\quad\quad {\rm for}\quad y>0,
\end{equation}
where $y$ denotes the exceedance over the threshold $t$. It has been shown \citep{Pickands1975} that, for very high thresholds, if the block maxima have an approximative distribution $F_{\rm GEV}\left(x;\alpha,\beta,\gamma\right)$, the distribution of the exceedances can be approximated by the \textit{generalized Pareto distribution} (GPD), given by
\begin{equation}\label{eq:GPD}
  F_{\rm GPD}\left(y;\tilde{\beta},\kappa\right) = \left\{ 
  \begin{array}{l l}
    1-\left[1+ \kappa\frac{y}{\tilde{\beta}}\right]^{-1/\kappa}, & \quad {\rm for}\quad\kappa\neq 0,\\
    1- \e^{-y/ \tilde{\beta}}, & \quad {\rm for}\quad\kappa = 0,\\
  \end{array} \right.
\end{equation}
where the GPD parameters are related to the GEV parameters via 
\begin{align}
\kappa &= \gamma,\\
\tilde{\beta} &=\beta + \gamma (t-\alpha),  \label{eq:betatilde}
\end{align}
and the exceedance $y$ is given by
\begin{equation}
y=x-t\quad {\rm if}\; x>t.
\end{equation}
Here all the parameters $\gamma$, $\alpha$ and $\beta$ are identical to the GEV parameters introduced before. For $t=\alpha$ the GPD and the GEV distribution are related by
\begin{equation}\label{eq:gpd_gev_relation}
F_{\rm GPD}\left(y;\tilde{\beta},\kappa\right)=1+\ln F_{\rm GEV}\left(x;\alpha,\beta,\gamma\right),
\end{equation} 
if $\ln F_{\rm GEV}\left(x;\alpha,\beta,\gamma\right) > -1$. In this way, once the GEV parameters from equation~\eqref{eq:parameters} are determined, also the GPD is fully determined. For small existence probabilities ($F_{\rm GEV}\left(x;\alpha,\beta,\gamma\right)$ close to $1$), both distributions will coincide according to equation~\eqref{eq:gpd_gev_relation}. This is shown in Fig.~\ref{fig:gpdvsgev}, from which it can be inferred that for clusters with an existence probability of less than $\sim 10$ per cent the two distributions differ by less than $1$ per cent. For galaxy clusters that are more likely to be found, the CDFs of both distribution start to significantly deviate from each other, such that, for a cluster with a $ \sim 50 $ per cent existence probability, the deviation is larger than $ 25 $ per cent. It is important at this point to note that both CDFs are correct in the sense that they just give answers to different statistical questions. The GEV distribution is the distribution of the most massive cluster to be found in a given cosmic volume, whereas the GPD is the distribution of the exceedance of all clusters above a high-mass threshold under the condition that the threshold is exceeded. In the case of rare clusters both distributions give identical answers, but for less rare clusters they can significantly differ.

Usually, when dealing with data based on an unknown underlying distribution and if the GPD parameters have to be determined from the data, then an appropriate choice of the threshold becomes crucial. When chosen too low, the limit law of the GPD is violated resulting in a bias; if it is chosen too high, data are so sparse that the variance in the parameter estimation is large. Since we will determine the GPD parameters via the formalism introduced in Sect.~\ref{sec:GEV},  we will study thresholds in the vicinity or above of the peak of the PDF inferred from the GEV. In this way, the limit law of the GPD can be assumed to be valid. This choice of the threshold is conservative and we do not attempt in this work to estimate the lowest possible threshold, which is usually done in an empirical way from the data. In principle there are two ways for the threshold estimation (see e.g. \cite{Coles2001}): the first one is based on the notion that the mean excess should approximately be a linear function of $t$ if the limit law is fulfilled and the second method is based on the expected stability of the estimated shape parameter for high enough thresholds. To assess the lowest possible choice of $t$ one would need to perform the analysis on a numerically simulated light cone, which we intend to do in a further study.
%-------------------------------------
%-------------------------------------
%-------------------------------------
\begin{figure*}
\centering
\includegraphics[width=0.33\linewidth]{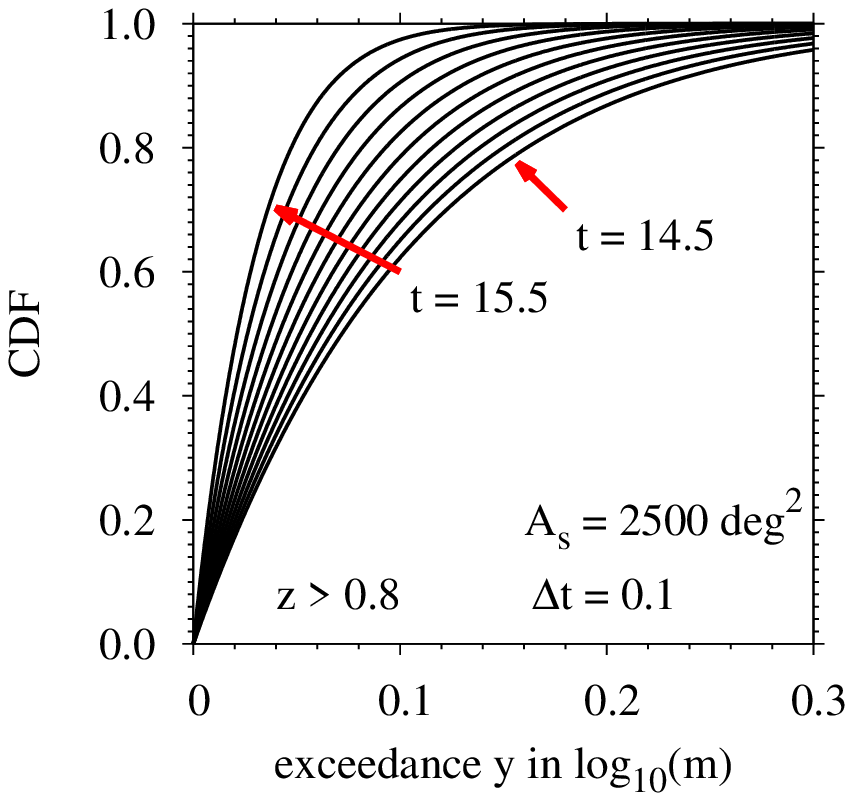}
\includegraphics[width=0.33\linewidth]{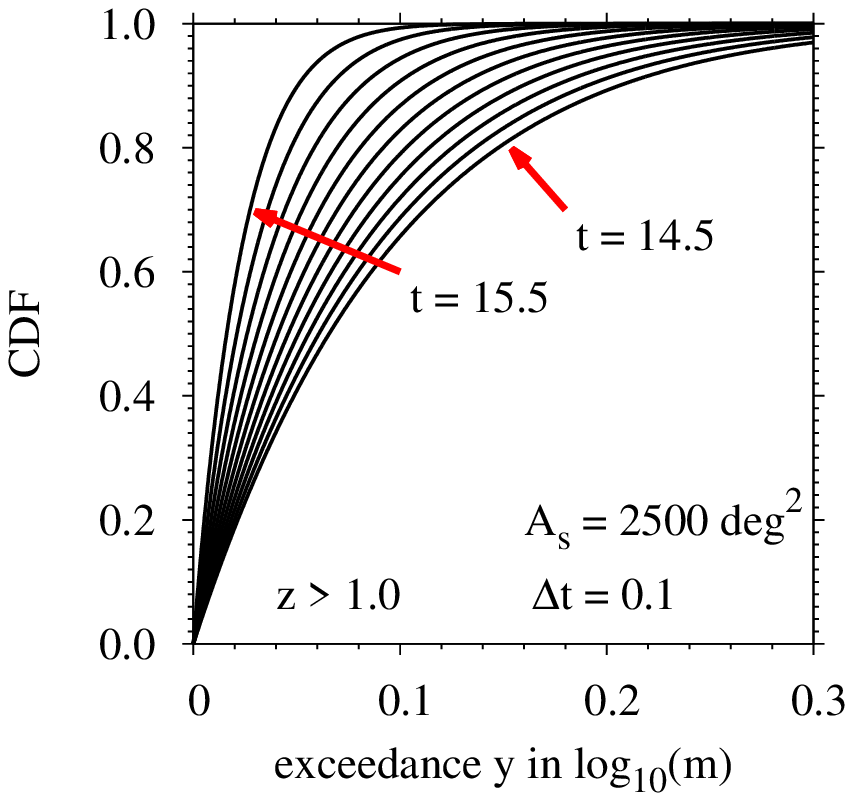}
\includegraphics[width=0.33\linewidth]{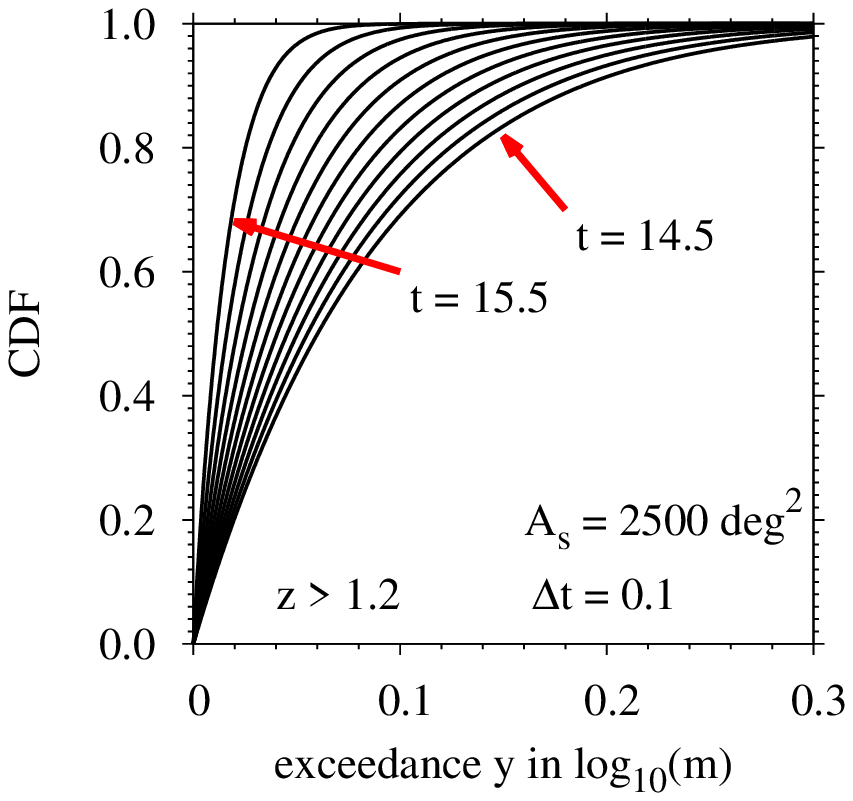}\\
\includegraphics[width=0.33\linewidth]{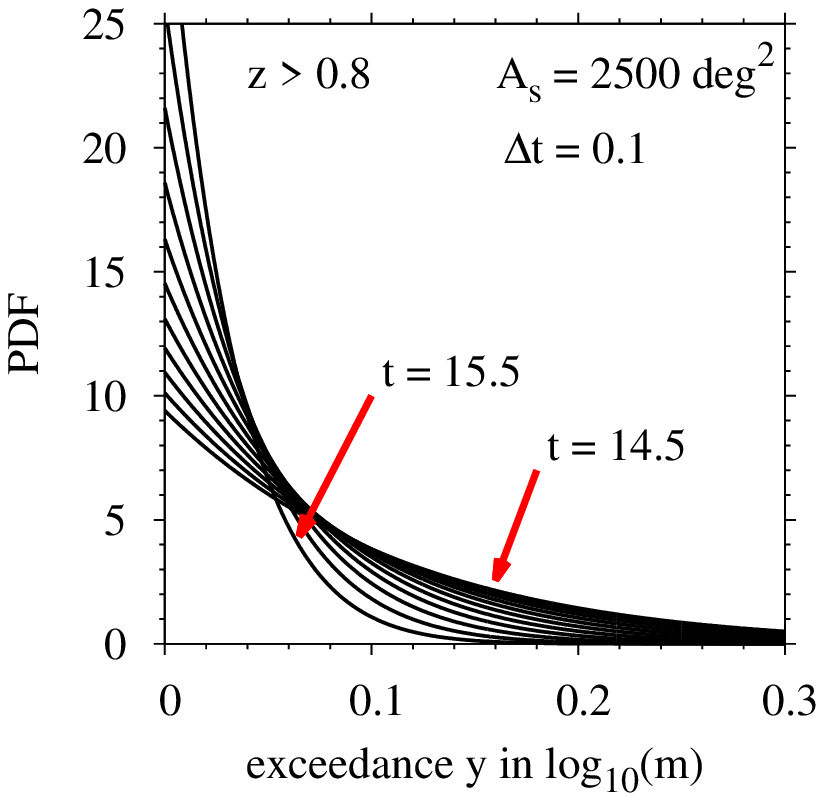}
\includegraphics[width=0.33\linewidth]{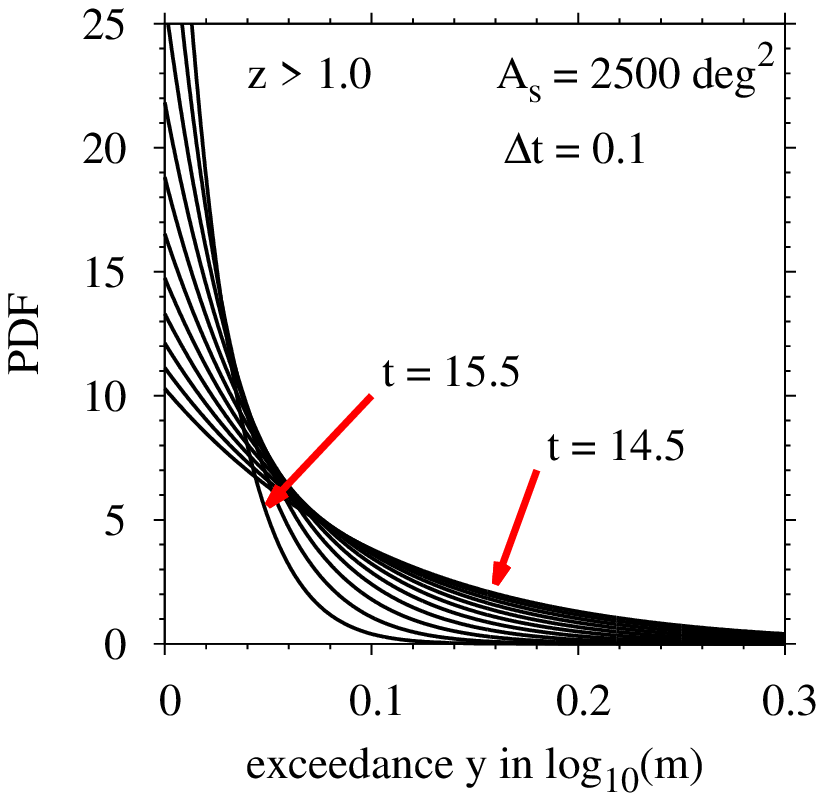}
\includegraphics[width=0.33\linewidth]{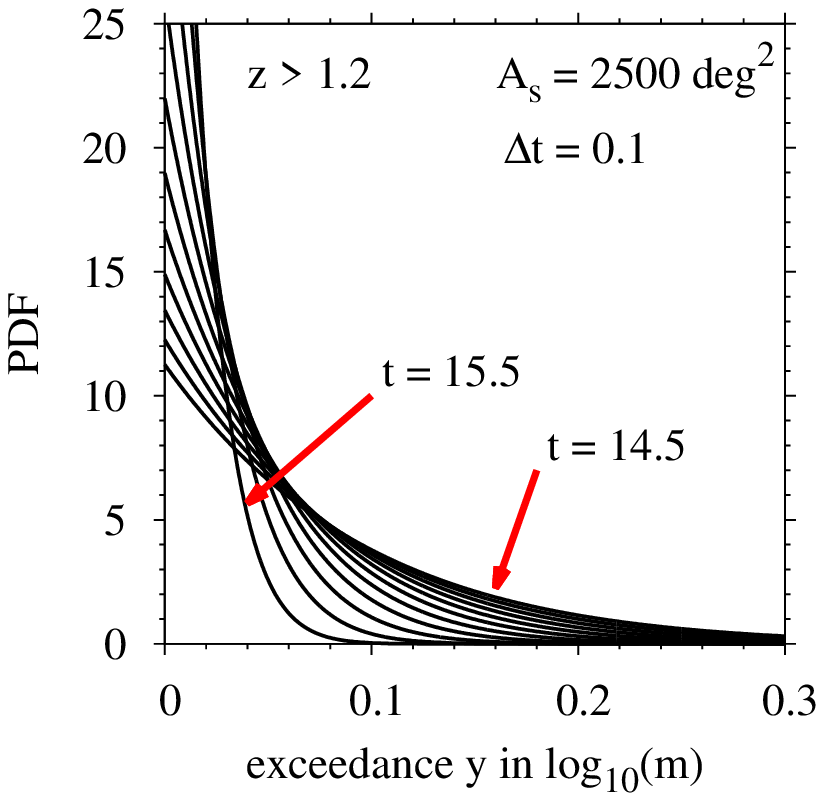}
\caption{CDFs (upper panels) and PDFs (lower panels) of the exceedance $y$ for a survey area $A_{\rm s} = 2\,500\,{\rm deg}^2$ and three different lower redshift limits in the range $0.8\le z\le 1.2$ as indicated in the individual panels. Each black line corresponds to a different value of the threshold $t$ that differ from each other by $\Delta t=0.1$. The outer values are indicated by the red arrows.}\label{fig:gpds}
\end{figure*}
%-------------------------------------
%-------------------------------------
\begin{figure}
\centering
\includegraphics[width=0.95\linewidth]{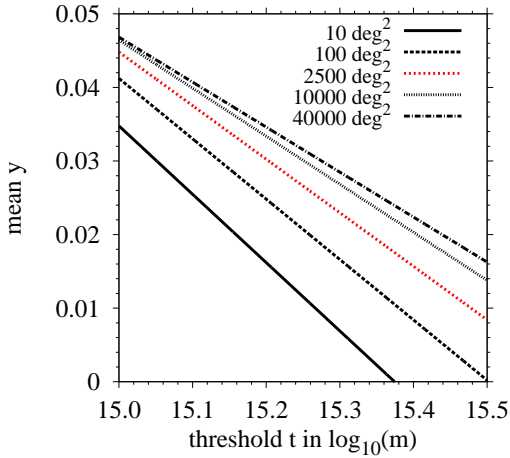}
\caption{Mean exceedance $y$ in the redshift interval $ 1.0\le z\le 1.5 $ as a function of the threshold $ t $ for different survey areas comprising $ 10\,{\rm deg}^2$, $ 100\,{\rm deg}^2$, $ 2\,500\,{\rm deg}^2$, $ 10\,000\,{\rm deg}^2$ and $ 40\,000\,{\rm deg}^2$, as labelled in the panel.}\label{fig:gpd_mean}
\end{figure}
%-------------------------------------
%------------------------------------------------------
\section{Modelling exceedances}\label{sec:exceed}
%------------------------------------------------------
It is very difficult to test cosmological models based on a single survey patch and thus on a single observation of the most massive cluster in the survey area. One solution to this problem is either to use many patches and to try to reconstruct the CDF of the most massive systems for a given patch-size and redshift range, or to take a single patch, provided the survey is deep enough, and to apply the exceedance approach introduced in Sect.~\ref{sec:GPD}. In doing so, instead of dividing the survey area into small patches, one uses all the information from objects above a given mass threshold, as illustrated in Fig.~\ref{fig:scheme}. At this point the survey selection function comes into play, which can cause a big problem to this approach since the limiting mass is usually an increasing function with redshift, which makes it difficult to define a threshold above which all clusters can be detected. An exception to this are surveys based on the SZ effect \citep{Sunyaev1972, Sunyaev1980}, since those exhibit an almost constant limiting survey-mass (see e.g. \cite{Carlstrom2002}) which would make them in principle ideally suited to an application of the exceedance theory.

As a case study we decided to consider a \textit{SPT}-like set-up with $A_{\rm s}=2\,500\,{\rm deg}^2$ \citep{Carlstrom2011} and different lower redshift limits in the range of $0.8\le z\le 1.2$. We are modelling thresholds $t=\log_{10}(m_{\rm threshold})$, where $m_{\rm threshold}$ is given in $M_\odot$, with $14.5\le t\le 15.5$ and compute the GPD distributions of the exceedances above a given threshold from the individual GEV analysis, as discussed in Sect.~\ref{sec:GEV}. The resulting CDFs and PDFs of the exceedances are presented in Fig.~\ref{fig:gpds}, where the former are shown in the upper panels and the latter in the lower panels. The start and end values for $t$ are indicated by the red arrows and the step-size used between each black curve is $\Delta t=0.1$. Both the CDFs and the PDFs show, as expected, that the higher the threshold is the less probable high exceedances are. Similarly the same holds for increasing the lower redshift, as can be seen by inspecting the panels from left to right.\\
Apart from the exceedance distributions themselves, it is also possible to calculate the mean exceedance $ E $ and the variance $S^2$, which are simply given by
\begin{eqnarray}
E    &=&\frac{\tilde\beta}{1-\gamma},\\
S^2&=&\frac{\tilde\beta^2}{(1-\gamma)^2(1-2\gamma)}.
\end{eqnarray}
For both relations the choice of the threshold enters via $\tilde{\beta}$, defined in equation~\eqref{eq:betatilde}, and the second moment exists only if the condition $\gamma < 1/2$ is fulfilled. This requirement is met for all cases of interest of this work. The dependence of the mean exceedance, $ E $, on the choice of the threshold is shown in Fig.~\ref{fig:gpd_mean} for five different choices of the survey area, $A_{\rm s}$, between $ 10\,{\rm deg}^2$ and $ 40\,000\,{\rm deg}^2$ and a redshift range of $1.0\le z\le 1.5$, which will be populated by future cluster surveys like \textit{EUCLID} \citep{Laureijs2011} for instance. The red dotted line illustrates the mean exceedance for the \textit{SPT} survey area of $2\,500\,{\rm deg}^2$. As expected, the mean exceedance is a decreasing function of the threshold, since the larger the threshold is the smaller the expected exceedances are. Of course, for a fixed threshold the larger survey area yields larger exceedances. Like the mean also the variance is a decreasing function of the threshold.
%-------------------------------------
\begin{figure*}
\centering
\includegraphics[width=0.325\linewidth]{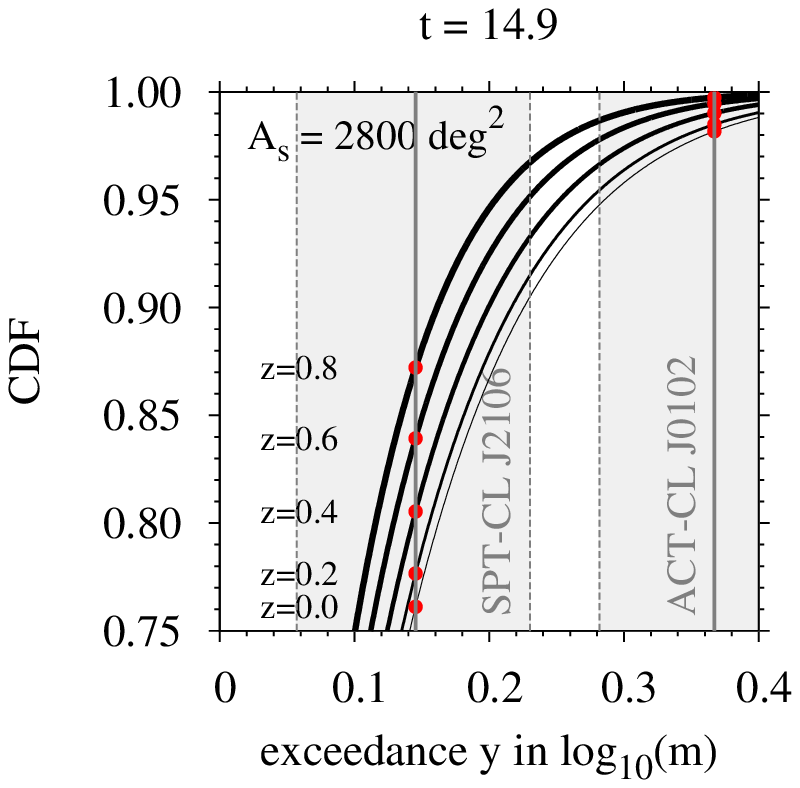}
\includegraphics[width=0.325\linewidth]{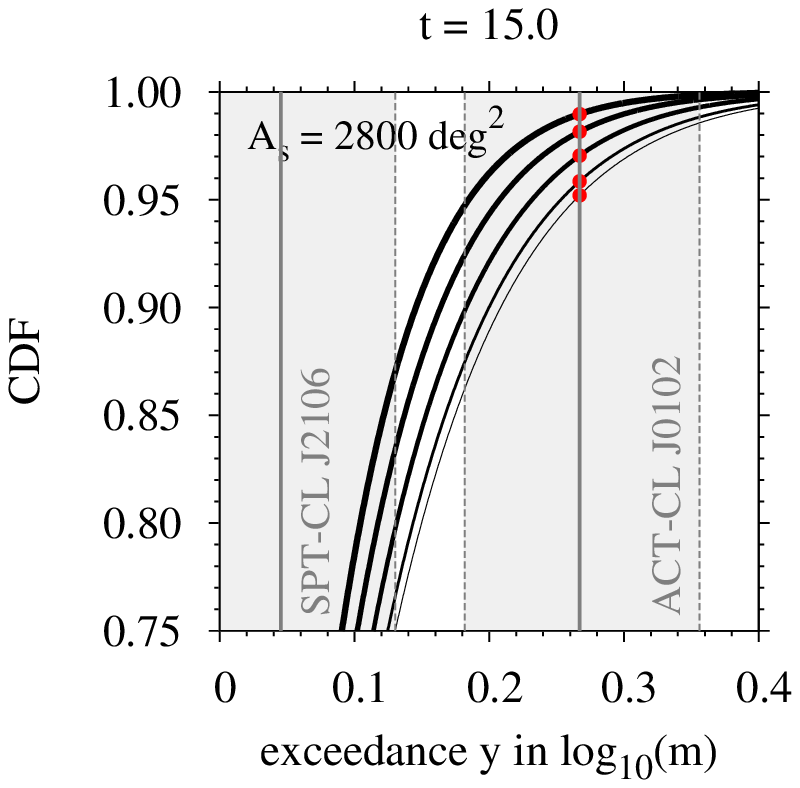}
\includegraphics[width=0.325\linewidth]{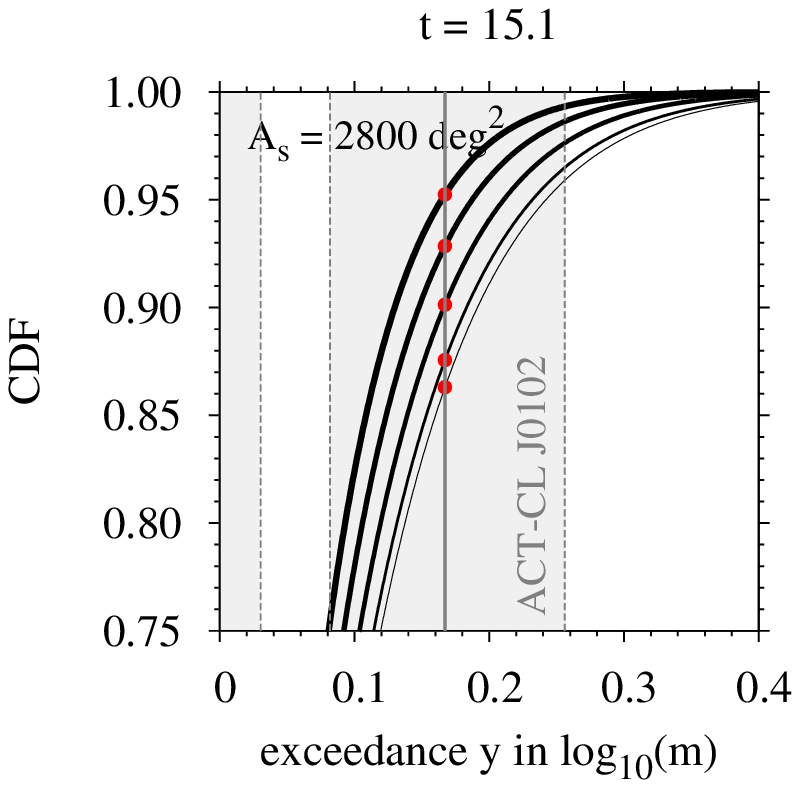}\\
\includegraphics[width=0.325\linewidth]{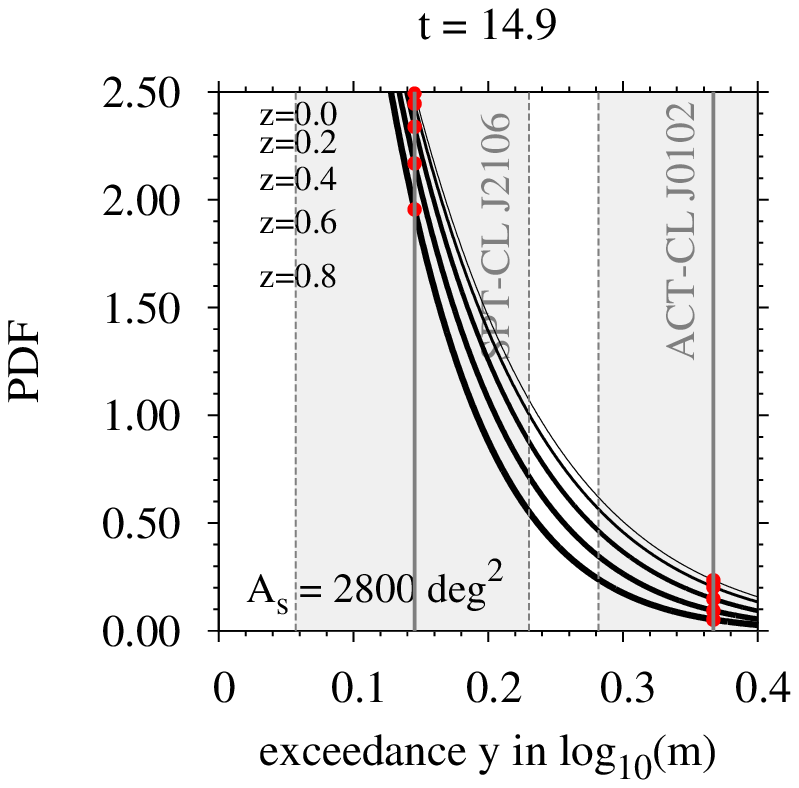}
\includegraphics[width=0.325\linewidth]{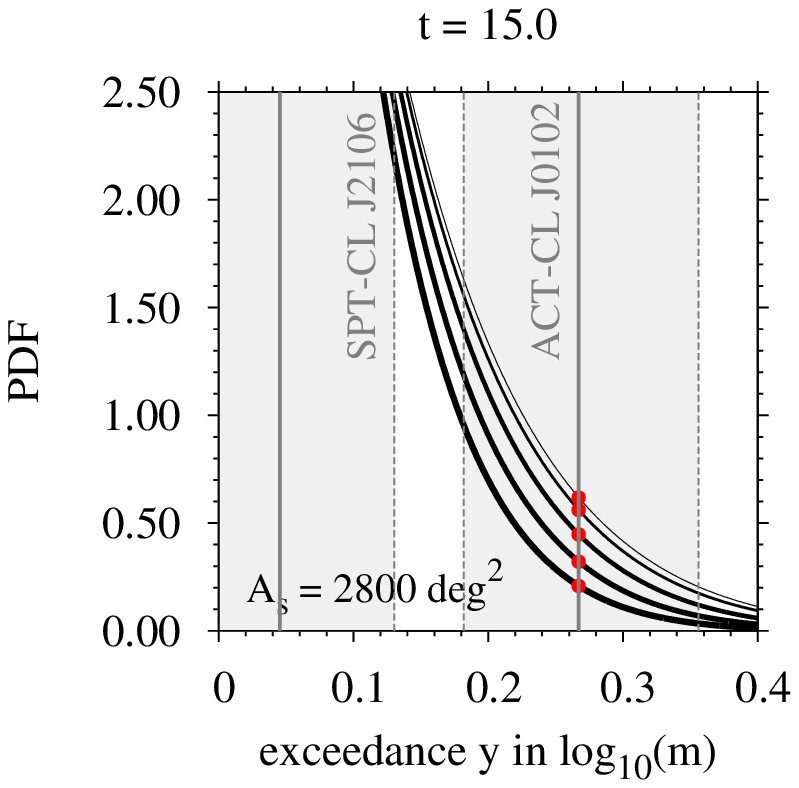}
\includegraphics[width=0.325\linewidth]{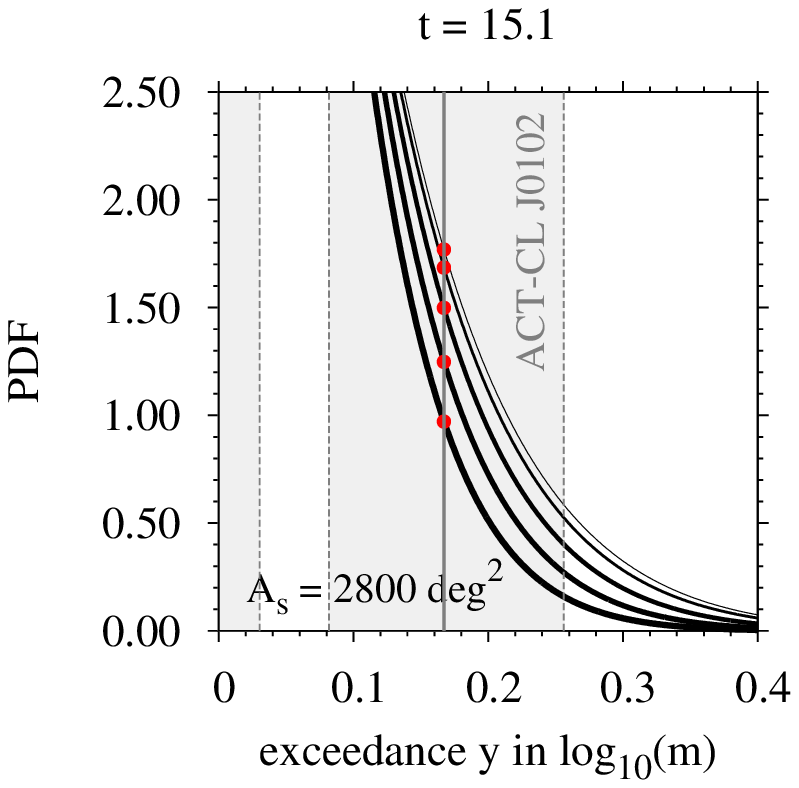}
\caption{CDFs (upper panels) and PDFs (lower panels) of the exceedance $y$ for a survey area $A_{\rm s} = 2\,800\,{\rm deg}^2$ and for five lower redshift limits in $z\in[0,0.8]$ (the upper redshift limit is kept constant at $z=3$), as indicated in the leftmost panels. The lines thicken with increasing lower redshift limit and the grey shaded areas show the allowed mass range due to uncertainties in the mass determination for the two most massive clusters in the combined \textit{ACT} and \textit{SPT} surveys. The red circles denote the different values of the CDFs and PDFs for the two clusters based on the different lower redshift limits.}\label{fig:single_clust}
\end{figure*}
%-------------------------------------
%------------------------------------------------------
\section{An example application: single clusters in the ACT \& SPT field} \label{sec:example_appl}
%------------------------------------------------------
After having introduced the basic theory in the previous sections, we present now an example application of the exceedance theory to two single SZ clusters in the \textit{Atacama Cosmology Telescope} (\textit{ACT}) \citep{Fowler2007} and \textit{SPT} \citep{Carlstrom2011} fields. We decided for two very massive objects, namely ACT-CL J0102 and SPT-CL J2106. The former one, also dubbed \textit{"El Gordo"}, has recently been discovered \citep{Marriage2011} by the \textit{ACT} in its $755\,{\rm deg}^2$ field. This merging system is currently the most massive cluster observed at $z>0.6$ \citep{Menanteau2011}. By combining SZ, optical , X-ray and infrared data, the mass could be determined to be $M_{200\rm m}=(2.16\pm 0.32)\times 10^{15}\,M_\odot$ at a spectroscopic redshift of $z=0.87$. Due to the fact that ACT-CL J0102 lies in the overlap region of the \textit{ACT} and \textit{SPT} survey areas, we conservatively assign the combined survey area of $2\,800\,{\rm deg}^2$ to this system.\\
The second object, SPT-CL J2106, has been detected by the \textit{SPT} collaboration \citep{Foley2011, Williamson2011} in a survey area of $2\,500\,{\rm deg}^2$. With a mass estimate of $M_{200\rm m}=(1.27\pm 0.21)\times 10^{15}\,M_\odot$ at a spectroscopic redshift of $z=1.132$, this extraordinary system is the most massive cluster at redshifts $z>1$.
%------------------------------------------------------
\subsection{Preliminary considerations} \label{subsec:prep_consid}
%------------------------------------------------------
Before performing a statistical analysis of single galaxy clusters, one has in general to account for two different effects that can substantially change the results.
\begin{enumerate}
\item The correction for the Eddington bias \citep{Eddington1913}: due to the steepness of the mass function at the high-mass end, it is more likely that lower mass systems scatter up than higher mass systems scatter down, resulting in a systematic shift to higher masses. Due to this effect, clusters appear to be rarer than they actually are.
\item The bias discussed in \cite{Hotchkiss2011} that stems from the a posteriori choice of the redshift interval for the statistical analysis. If the lower redshift boundary is set to the cluster redshift, one pushes the rareness of a given cluster to the maximum. However, a cluster of a given mass could have easily shown up at another redshift. If not accounted for, this bias, like the Eddington bias, lets clusters appear to be rarer than they are.  
\end{enumerate} 
The strategies for correcting for these effects have been already discussed in \cite{Waizmann2011b}, thus we will only briefly summarise them at this point. We correct for the Eddington bias, following \cite{Mortonson2011}, by shifting the observed mass, $M_{\rm obs}$, to a corrected mass, $M_{\rm corr}$, by
\begin{equation}
\ln M_{\rm corr}=\ln M_{\rm obs}+\frac{1}{2}\epsilon\sigma_{\ln M}^2,
\end{equation} 
where $\epsilon$ is the local slope of the mass function ($\dd n/\dd \ln M\propto M^\epsilon$) and $\sigma_{\ln M}$ is the uncertainty in the mass measurement (for more details see \cite{Waizmann2011b}). We account for the bias discussed in \cite{Hotchkiss2011} by a priori choosing the redshift ranges $z\in[z_{\rm low},z_{\rm up}]$. In order to compare theory with observations on the same grounds, it is usually necessary to unify the mass definitions. The observationally reported masses are frequently defined considering an overdensity computed with respect to the critical one $(M_{200\rm c})$, whereas in the mass function of \cite{Tinker2008} for instance, the mean background density $(M_{200\rm m}$) is assumed. For the clusters we discuss in this work, no correction is necessary because the observed masses are already given in $M_{200\rm m}$. Therefore, we adopt from \cite{Waizmann2011b} for the mass of ACT-CL J0102 a value of $M_{200\rm m}^{\rm Edd}=1.85_{-0.33}^{+0.42}\times 10^{15}\,M_\odot$ and for SPT-CL J2106 a value of $M_{200\rm m}^{\rm Edd}=1.11_{-0.20}^{+0.24}\times 10^{15}\,M_\odot$ that we will use hereafter.
%------------------------------------------------------
\subsection{Results}  \label{subsec:single_results}
%------------------------------------------------------
The results of our GPD analysis are shown in Fig.~\ref{fig:single_clust}, in which we present the CDFs (upper panels) and PDFs (lower panels) of the exceedance over different thresholds $t\in \lbrace 14.9, 15.0,15.1\rbrace$ (from left to right) based on a combined survey area of $2\,800\,{\rm deg}^2$ and for different lower redshift limits comprising $z_{\rm low}\in\lbrace0, 0.2, 0.4, 0.6, 0.8\rbrace$. The upper redshift limit is kept fixed at $z_{\rm up}=3$, since it has only a weak impact on the distribution functions. The grey shaded areas denote the uncertainties in the observed masses. As expected, ACT-CL J0102 sits further in the tail of the distributions than SPT-CL J2106, because its mass is higher and thus the exceedance is larger. With increasing the threshold the clusters move to smaller exceedances and become thus more likely to be found. As discussed in \cite{Hotchkiss2011}, the distributions are sensitive to the choice of $z_{\rm low}$ in the sense that for smaller $z_{\rm low}$ the clusters are more likely to be found.

\begin{figure*}
\centering
\includegraphics[width=0.495\linewidth]{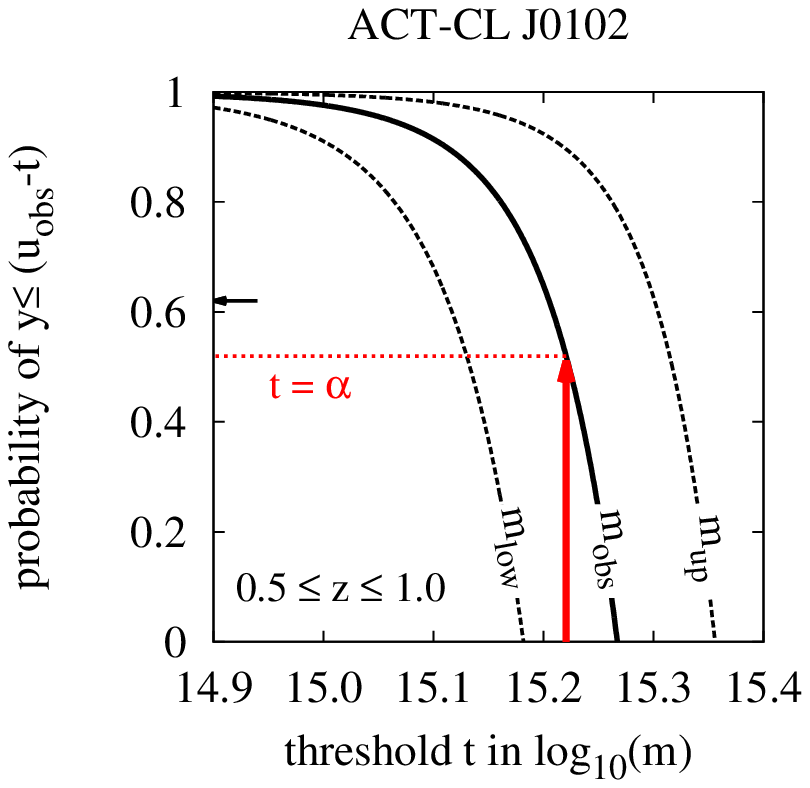}
\includegraphics[width=0.495\linewidth]{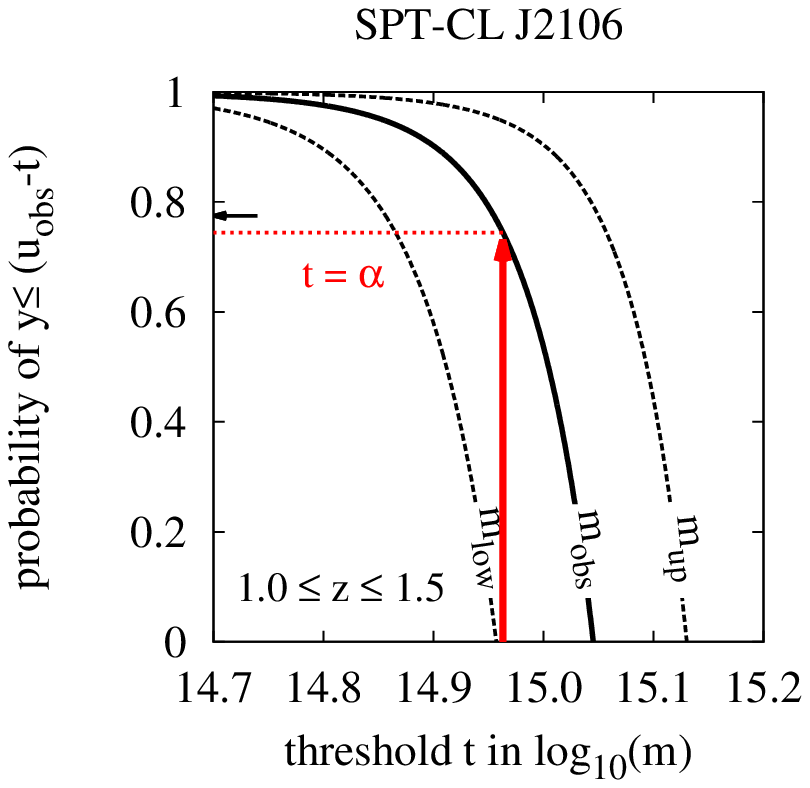}
\caption{CDFs as a function of the threshold $t$ for ACT-CL J0102 (left-hand panel) and for SPT-CL J2106 (right-hand panel) for fixed observed masses $u_{\rm obs}\equiv \log_{10}m_{\rm obs}$ in their respective survey areas of $A_{\rm s} = 2\,800\,{\rm deg}^2$ and $A_{\rm s} = 2\,500\,{\rm deg}^2$. The dashed lines denote the uncertainties in the measured masses. The corresponding redshift intervals are given in the left, lower corner of each panel. The red, vertical, arrows denote the results for the particular choice of $t=\alpha$ and the small black, horizontal arrows denote the result of a full GEV analysis.}\label{fig:varying_t}
\end{figure*}

We also calculated the probability to find $y\le (m_{\rm obs}-t)$ for a fixed observed mass, $m_{\rm obs}$, as a function of threshold and present the results in Fig.~\ref{fig:varying_t} for ACT-CL J0102 (left panel) and for SPT-CL J2106 (right panel). For the former system, we use $A_{\rm s} = 2\,800\,{\rm deg}^2$ and the redshift interval $0.5\le z\le 1.0$. For the latter, we use $A_{\rm s} = 2\,500\,{\rm deg}^2$ and the redshift interval $1.0\le z\le 1.5$. We chose these specific redshift intervals a priori in order to avoid the previously mentioned bias and to be directly comparable with the study in \cite{Waizmann2011b}. The black solid line denotes $m_{\rm obs}=M_{200\rm m}^{\rm Edd}$ and the black dashed lines denoted the upper (lower) allowed mass limits $m_{\rm up}$ ($m_{\rm low}$). We also added a red arrow together with a red dotted line in order to denote the probability for the particular choice of the threshold, $t=\alpha$, and for comparison we added a small black arrow pointing to the probability obtained from a GEV analysis \citep{Waizmann2011b}.
From the position of the black arrow with respect to the red dotted line, one can infer that the GPD delivers lower probabilities of existence than the GEV analysis. The reason and nature of this difference has already been discussed in Section~\ref{sec:GPD}. As shown in Fig.~\ref{fig:gpdvsgev}, the two approaches give similar results for very rare clusters only. This is also the reason why the difference between the arrow and the red dotted line is smaller for SPT-CL J2106 in the right-hand panel compared to  ACT-CL J0102 in the left-hand panel. Since the redshift intervals have been chosen a priori, the former cluster appears to be rarer than the latter. At this point it should be repeated that both probabilities from the GEV and the GPD approach are correct, since they are not the answer to the same statistical question. 
%-------------------------------------
%-------------------------------------
%------------------------------------------------------
\section{On the applicability of the GPD to parameter estimation} \label{sec:param_estim}
%------------------------------------------------------
Apart from theoretically modelling exceedances for a given cosmological model, the GPD-based approach could be advantageous for the estimation of the GEV parameters $\alpha$, $\beta$ and $\gamma$. Instead of only using block-maxima (the most massive clusters observed in smaller patches) as suggested in \cite{Waizmann2011a}, a GPD approach would use the information from all observed systems above a given high-mass threshold. This difference could be particularly important for mass-limited SZ surveys in the sense that information of a larger number of objects could be used for the parameter estimation. In the following, we will study in more detail the performance of the GPD approach for parameter estimation and eventually its usability as a cosmological probe.

\begin{figure*}
\centering
\includegraphics[width=0.325\linewidth]{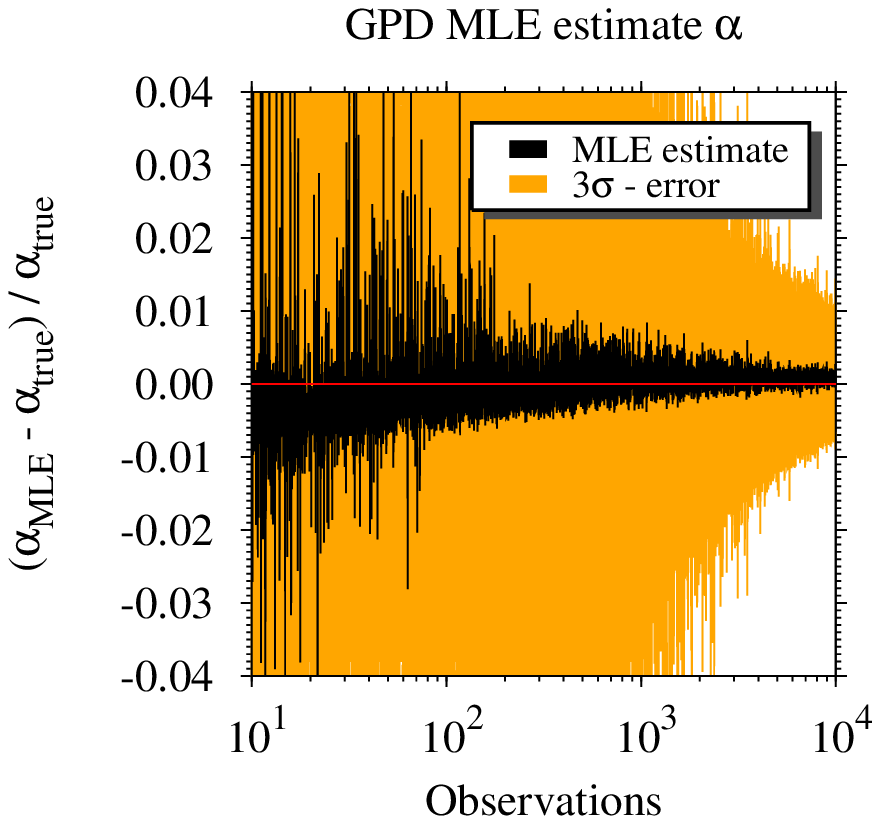}
\includegraphics[width=0.325\linewidth]{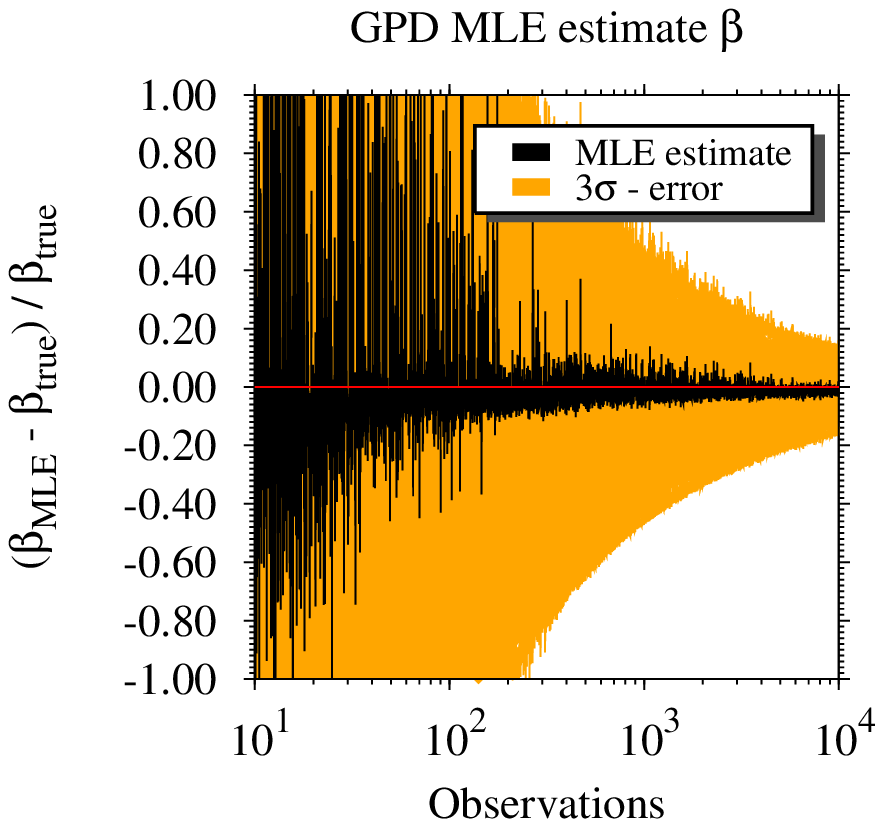}
\includegraphics[width=0.325\linewidth]{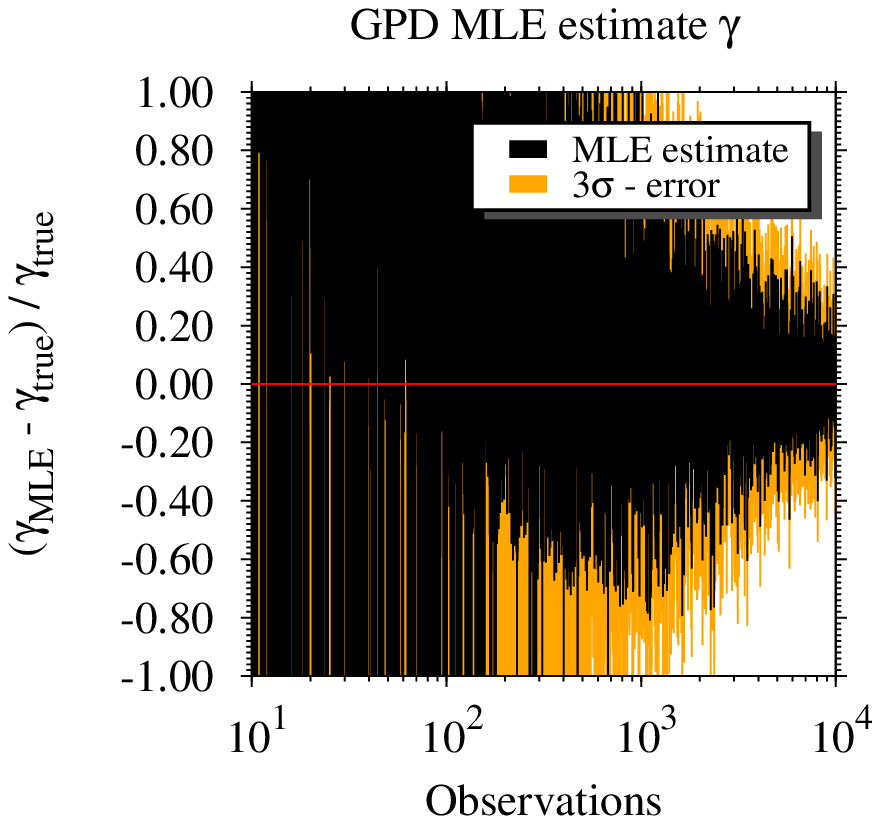}\\
\includegraphics[width=0.325\linewidth]{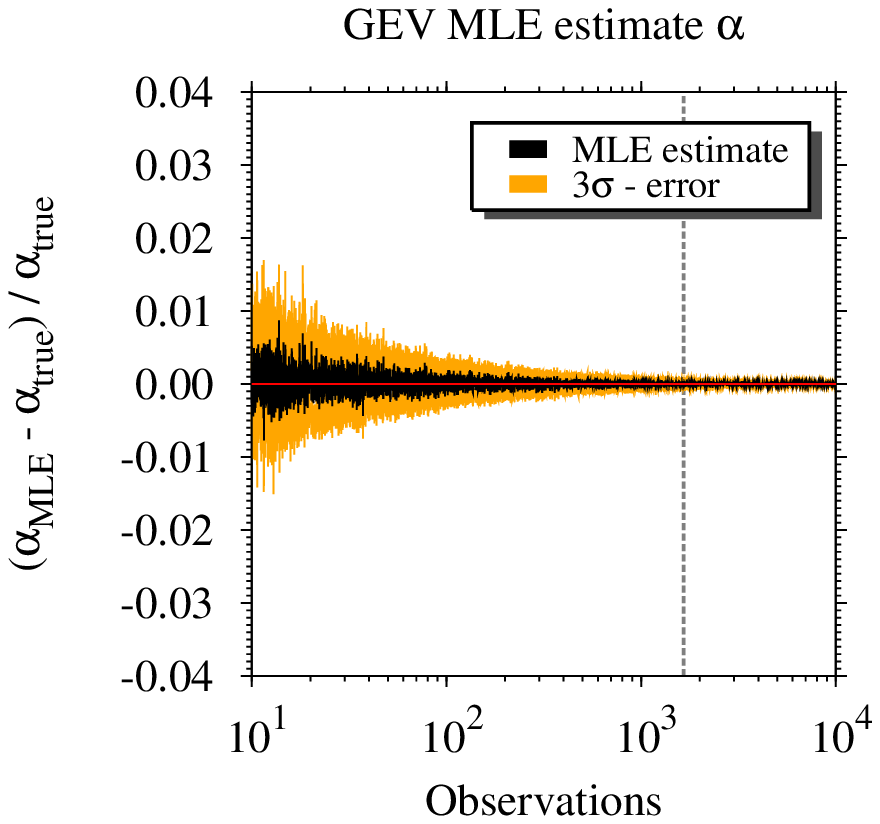}
\includegraphics[width=0.325\linewidth]{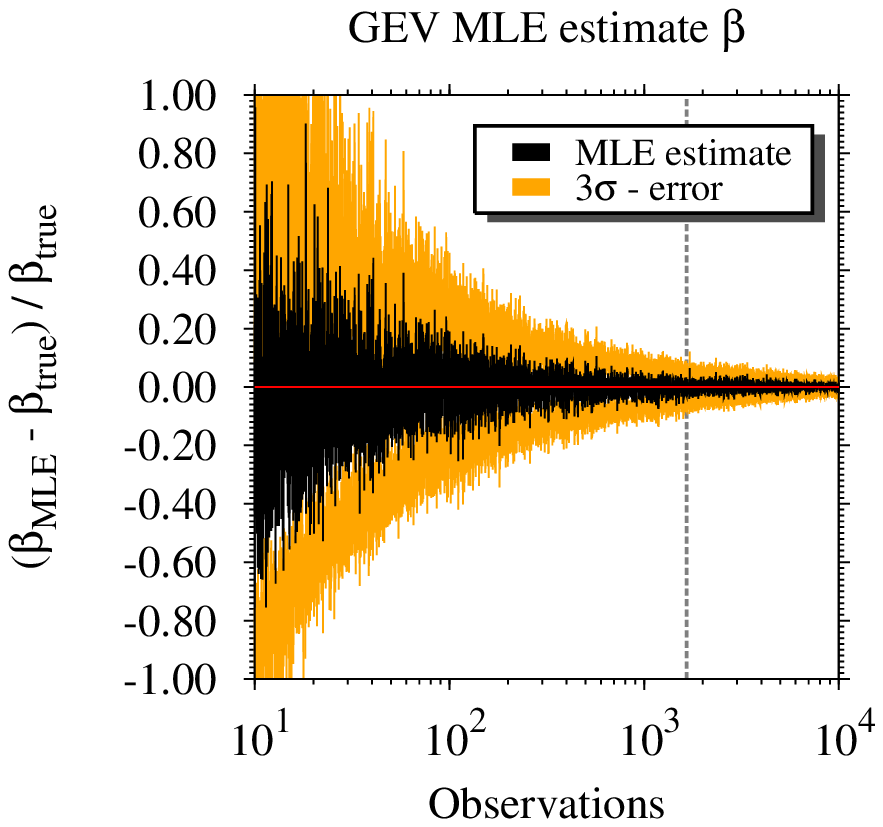}
\includegraphics[width=0.325\linewidth]{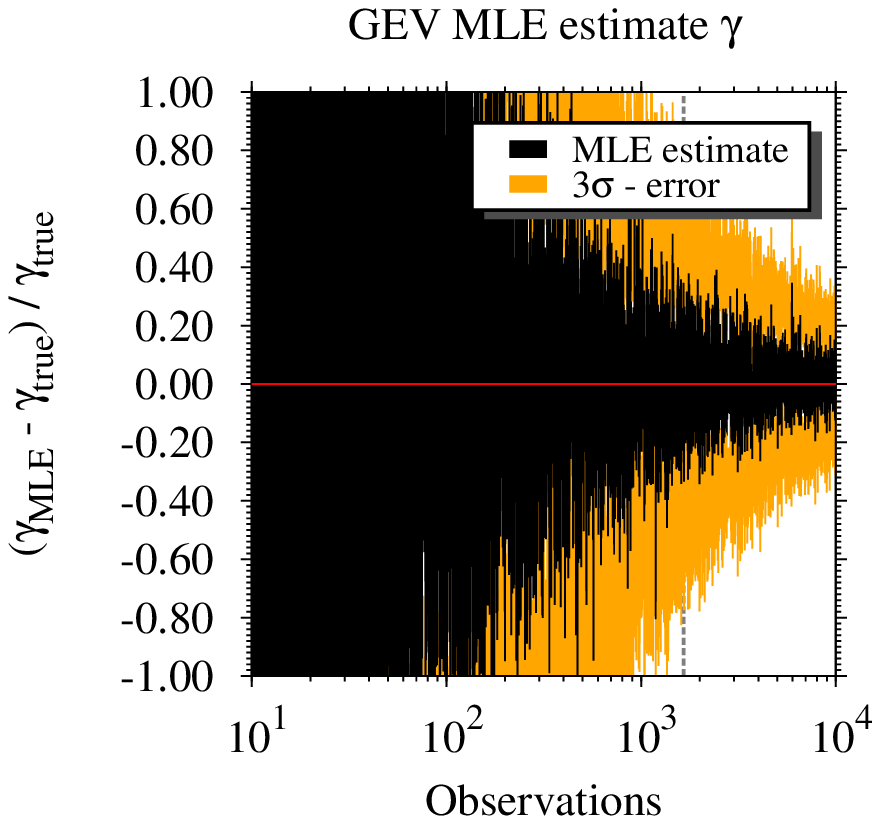}
\caption{Relative differences between the MLE estimates $\alpha_{\rm MLE}$, $\beta_{\rm MLE}$ and $\gamma_{\rm MLE}$ (in the order of the panels) and the true underlying parameters $\alpha_{\rm true}$, $\beta_{\rm true}$ and $\gamma_{\rm true}$ as a function of the number of observations for the GPD-based modelling of exceedances over the threshold $t=15.0$ (upper panels) and for the GEV-based modelling of the CDF of the most massive clusters in patches (lower panels). The black lines denote the MLE estimates and the orange area shows the $3\sigma$ errors. The survey area was assumed to be $A_{\rm s}=2\,500\,{\rm deg}^2$ for the GPD case and the patch size was assumed to be $A_{\rm p}=25\,{\rm deg}^2$ for the GEV case. The redshift range is in both cases $0.5\le z\le 3.0$ and in the lower panels the vertical dotted line indicates what number of observations (patches) corresponds to the full sky.}\label{fig:GPD_GEV_MLE}
\end{figure*}

For the estimation of the distribution parameters of the GPD, we utilise the maximum likelihood estimation (MLE) method. The log-likelihood function for the observation of $n$ excesses over the threshold $t$ reads
\begin{equation}
\ln\mathcal{L}=\left\{ 
  \begin{array}{l l}
  -n\ln\tilde\beta - \left(1+\frac{1}{\gamma} \right)\sum_{i=1}^n\ln\left(1+\frac{\gamma y_i}{\tilde{\beta}}\right), & \quad {\rm for}\quad\gamma\neq 0,\\
      -n\ln\tilde\beta - \frac{1}{\tilde{\beta}}\sum_{i=1}^ny_i, & \quad {\rm for}\quad\gamma = 0,\\
  \end{array} \right.
\end{equation}
where $\tilde\beta$ from equation~\eqref{eq:betatilde} contains the parameter dependence on $\alpha$, $\beta$ and $\gamma$. For the best estimates of the GPD parameters, one minimizes $-\ln\mathcal{L}$ for the given set of $y_i$. For the numerical minimization process, we utilized the MINUIT2 library\footnote{http://www.cern.ch/minuit}. In the statistical literature it is very common to consider the GPD distribution as a 2-parameter distribution of $\gamma$ and $\tilde\beta$ (see e.g. \cite{Huesler2011}). However, as shown in \cite{Waizmann2011a}, the location parameter, $\alpha$, is, among the three distribution parameters, the one with the strongest dependence on the underlying cosmological model and, thus, it is not desirable to mask this parameter by combining it with the parameters $\gamma$ and $\beta$ to form the unified parameter $\tilde\beta$. Therefore, we will focus our study on the 3-parameter case.\\
In order to understand whether we can expect an improvement in the parameter estimation with GPD or not, we sample observations from the true calculated GPD distribution along the lines of Section~\ref{sec:GPD} and use these samples to estimate the parameters using MLE. The results of this procedure are shown in the upper panels of Fig.~\ref{fig:GPD_GEV_MLE} for $\alpha$, $\beta$ and $\gamma$ from left to right. We chose arbitrarily the redshift interval of $0.5\le z \le 3.0$, a threshold of $t=15.0$ and an \textit{SPT}-like survey area of $A_{\rm s}=2\,500\,{\rm deg}^2$. The threshold is chosen to be high enough to assume the validity of the limit law of the GPD and to mimic the set-up of the \textit{SPT} high-mass cluster sample \citep{Foley2011}. In all three panels we show the relative difference between the MLE estimated and the true underlying parameters as a function of the number of observations or, more clearly, the number of clusters above the threshold. The black lines show the parameter estimates and the orange area denotes the $3\sigma$ error range. From the two rightmost panels, it can directly be inferred that the scale parameter, $\beta$, and the shape parameter, $\gamma$, can, even for a fictitious large number of observations, not be reliably estimated. For the location parameter, the situation seems too be much better but, particularly for a small number of observations, the estimate seems to be biased low and moreover the $3\sigma$ error range is quite large. This first result is sobering considering that at first sight the GPD based approach seemed to be advantageous due to the increased amount of objects.

In order to compare the results of the MLE for the GPD with the performance of a pure GEV-based approach, we repeated the previous statistical experiment with a GEV distribution for the same redshift range. The log-likelihood function for the GEV case is given by
\begin{equation}
\ln\mathcal{L}=-n\ln\beta - \sum_{i=1}^n \left(1+\frac{1}{\gamma}\right)\ln\left(1+\gamma\frac{(u_i-\alpha)}{\beta}\right)+\left(1+\gamma\frac{(u_i-\alpha)}{\beta}\right)^{-1/\gamma},
\end{equation}
where $n$ is the number of observed clusters and $u_i=\log_{10}M_i$ are the individual observed masses. In order to mimic the different methodology of dividing the survey area into smaller patches, we divide the survey area, $A_{\rm s}$, into $n_{\rm p}$ equally sized patches of area $A_{\rm p}$. We fix $A_{\rm p}=25\,{\rm deg}^2$ such that in a \textit{SPT}-like survey one would observe $100$ patches. The results of this procedure are shown in the lower panels of Fig.~\ref{fig:GPD_GEV_MLE} again for all three GEV parameters $\alpha$, $\beta$ and $\gamma$. The difference in the performance of the parameter estimation with respect to the GPD approach is substantial. The statistical errors are much smaller for $\alpha$ and $\beta$. Particularly for the location parameter, $\alpha$, percent-level estimation in $100$ patches would be possible in this idealised case. The estimation of the scale, and especially of the shape parameter are less precise but still significantly better with respect to the GPD approach. Furthermore, the parameter estimates are unbiased for the idealised GEV case, even for a small number of observations.

\begin{figure}
\centering
\includegraphics[width=0.95\linewidth]{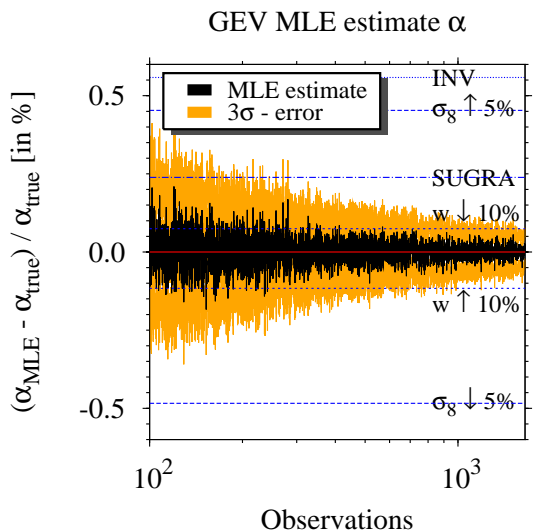}
\caption{Relative percent differences between $\alpha_{\rm MLE}$ and the true underlying parameters $\alpha_{\rm true}$ as a function of the number of observations for GEV-based modelling of the CDF of the most massive clusters in patches. In addition, we added the expected relative differences for a $5$ per cent increase (decrease) in $\sigma_8$ (blue dashed line), $10$ per cent increase (decrease) in $w$ (blue short-dashed line), the SUGRA (blue dashed dotted) and the INV model (blue dotted line). The black lines denote the MLE estimates and the orange area shows the $3\sigma$ errors. The patch size was assumed to be $A_{\rm p}=25\,{\rm deg}^2$, which corresponds to a division of the \textit{SPT} field into 100 patches and the redshift ranges from $1.0\le z\le 3.0$.}\label{fig:GEV_MLE_detail}
\end{figure}

In order to understand better how the achievable accuracy in the measurement of $\alpha$ compares to deviations from $\Lambda$CDM, we added in Fig.~\ref{fig:GEV_MLE_detail} the relative changes in the parameter $\alpha$ for variations of $\sigma_8=0.811$ by $\pm 5$ per cent and of the equation of state parameter $w=-1$ by $\pm 10$ per cent, keeping the other cosmological parameters fixed, respectively. We also added a quintessence model with an inverse power-law potential (INV) and a supergravity model (SUGRA), identical to the ones used in \cite{Pace2010}. Based on the results of \cite{Waizmann2011a}, according to which the GEV-based approach is particularly sensitive to deviations from the $\Lambda$CDM model at high redshifts, we choose a redshift range of $1.0\le z\le 3.0$. The patch size was assumed again to be $A_{\rm p}=25\,{\rm deg}^2$. It can be seen that even for $\sim 100$ patches the Gnedenko approach allows good constraints on $\sigma_8$, which will be of course degenerate with $\Omega_{{\rm m}0}$. The constraints on $w$ are less tight and would require the combination with other cosmological probes to constrain this parameter with a higher precision; for the INV and SUGRA models $300-400$ patches would be sufficient to rule them out with this strongly idealised statistical experiment.

With this small statistical experiment, we could show that the Pareto approach seems not to be a favourable approach to improve the GEV parameter estimates. On the contrary, our results confirm that the patch-based Gnedenko approach is by far superior for the estimation of the location parameter, $\alpha$, which is the most interesting parameter for cosmological applications. Of course, observational effects and biases will lower the performance in the estimation of $\alpha$, yet from a statistical point of view the method based on the Gnedenko approach seems to be favoured. Even for a small number of observations, the MLE estimates behave extremely well. With this solid statistical foundation the next step will be an application of the Gnedenko approach to real observables rather than cluster mass in order to examine how well the method performs when applied to real data.
%-------------------------------------
%-------------------------------------
%-------------------------------------
%-------------------------------------
%------------------------------------------------------
\section{Summary and conclusions}\label{sec:conclusions}
%------------------------------------------------------
In this work we have presented for the first time an application of the generalized Pareto distribution to model the exceedances of galaxy clusters over a high-mass threshold. The approach to model exceedances over high thresholds is very closely linked to the modelling of extreme values by means of the GEV distribution. Instead of calculating the distribution of the block maxima which correspond to the most massive galaxy cluster in a given cosmic volume, one models the distribution of all clusters that are found to be above a given high-mass threshold under the condition that this threshold is exceeded. We related the underlying cosmological model to the three GEV parameters and related those to the two GPD parameters that fully describe the distribution.

We showed that, for a particular choice of the threshold $ (t=\alpha) $, the CDFs of both the GPD and the GEV lead to basically identical values if the galaxy cluster is very rare (existence probability $ \lesssim 10 $ per cent). For clusters that are less rare, both CDFs quickly deviate substantially from each other. However, it is important to note that both distributions are correct in the sense that they are answers to different statistical questions: the GEV distribution is the distribution of the most massive cluster to be found in a given cosmic volume, whereas the GPD is the distribution of the exceedance of all clusters above a high-mass threshold under the condition that the threshold is exceeded.

Based on the argument that, in contrast to almost all other types of cluster surveys, SZ ones exhibit a constant limiting mass out to high redshifts, we study the application of GPD for an \textit{SPT}-like survey with a survey area of $2\,500\,{\rm deg}^2$. We calculate the probability distributions of the exceedances for a range of thresholds and redshift bins. As expected, the PDFs fall steeper to zero the larger the threshold is chosen for a fixed survey area and redshift range. The same applies if the threshold is kept fixed but one considers the volume of interest to be placed at higher redshifts. 

With the possibility to analytically derive the distribution of the threshold excesses, we apply the GPD approach to two SZ clusters, namely ACT-CL J0102 and SPT-CL J2106, that have been observed in the combined survey area of ACT and SPT. For our calculations we assigned to each individual system a priori a redshift range for which we perform the analysis. This is done in order to avoid the bias that arises from the a posteriori choice of the volume, as discussed in \cite{Hotchkiss2011}. None of the two clusters is in tension with the $ \Lambda $CDM cosmology, as discussed in \cite{Waizmann2011b}, and also the GPD-based approach leads to the same conclusion. This result is in agreement with the conclusions drawn in the recent works of \cite{Chongchitnan2011}, \cite{Harrison&Coles2011b} and \cite{Menanteau2011} who find no tension with $\Lambda$CDM for individual clusters. Since SPT-CL J2106, due to its position in the redshift interval, appears to be rarer than ACT-CL J0102, the GEV and GPD approach deliver very similar probabilities, as it has been discussed above.

So far we summarized the results for the potential application to single systems. In the second part of this work, we discussed whether the GPD might potentially be used as a cosmological probe or not. In \cite{Waizmann2011a}, we proposed to utilise GEV as a cosmological probe by means of dividing the survey area in equally sized patches and to measure the mass of the most massive cluster in the patch. In this way it would be possible to reconstruct the distribution of the maxima and to compare it with the theoretical expectations. We could show that the position of the peak of the PDF is the most promising GEV parameter due to its strong dependence on the underlying cosmology.

For a survey like the \textit{SPT} one, the determination of the distribution parameters via maximum likelihood methods should be superior in the GPD case with respect to the GEV one, since we expect more clusters to be found above a threshold than block maxima by dividing the survey into smaller patches. The increased amount of information should in principle reduce the variance in the maximum likelihood estimates of the parameters and therefore result in tighter constraints on deviations from the $\Lambda$CDM model. In order to test this naive assumption we performed a sampling experiment for which we created random realisations of the GPD and the GEV distributions and calculated the MLE-estimates for different sample sizes. We found that the location parameter $ \alpha $, which is tightly related to the most likely maximum mass that should be found in a given volume, can be estimated with the highest precision with respect to the two other GEV parameters $ \beta $ and $ \gamma $ in both approaches. However, in the GPD approach a much larger number of realisations (clusters) is needed with respect to the patch-based GEV approach. For the latter already $ \sim 100 $ patches are sufficient to reach a percent level accuracy on $ \alpha $. From this point of view, it seems that the GEV based approach is far superior to the GPD based one. The remaining challenge, however, will be to get observational biases stemming from uncertainties in the mass measurements and the resulting confusion of less massive clusters as the most massive one.

Thus, the main conclusions of this work can be summarized as follows.
\begin{enumerate}
\item The excess of very massive high-mass clusters can be analytically modelled with the generalized Pareto distribution.
\item For rare clusters, the GPD and the GEV based modelling lead to identical existence probabilities for extreme galaxy clusters.
\item Modelling of exceedances by means of the GPD approach seems to be disfavoured as a cosmological probe when compared to the patch-based GEV approach.
\item Utilising a MLE approach, the location parameter, $ \alpha $, can be estimated under idealised circumstances on the percent level for less than $ \sim 100 $ patches.
\end{enumerate}
The last point indicates that, from a statistical point of view, the patch-based method can be easily applied to relatively small survey areas, particularly if the focus lies on high-$ z $ systems. The GPD approach, however, seems only to be usefully applicable for studies of single objects but not as a cosmological probe. In order to observe the large number of clusters required to get an accurate estimate of $ \alpha $, the threshold would have to be substantially lowered and this would violate the assumption on which the GPD is based. In addition very large survey areas would be required as well, which makes the GEV based approach more appealing for a real application of the method.
%------------------------------------------------------
\section*{Acknowledgments}
%------------------------------------------------------
We acknowledge financial contributions from contracts ASI-INAF I/023/05/0, ASI-INAF I/088/06/0, ASI I/016/07/0 COFIS, ASI Euclid-DUNE I/064/08/0, ASI-Uni Bologna-Astronomy Dept. Euclid-NIS I/039/10/0, and PRIN~MIUR~2008 "Dark energy and cosmology with large galaxy surveys". 

\bibliographystyle{mn2e}

\label{lastpage}
   
\end{document}